\documentclass[twocolumn,superscriptaddress,preprintnumbers,amsmath,amssymb,floatfix]{revtex4-1}

\pdfoutput=1
\usepackage{graphicx} 
\usepackage{dcolumn} 
\usepackage{bm}
\usepackage{color}
\usepackage{hyperref}
\usepackage{graphicx}
\usepackage{amsmath}
\usepackage{setspace}

\begin{document}

\title{A local-density-approximation description of high-momentum tails in isospin asymmetric nuclei}
\author{Xiao-Hua Fan}
\affiliation{School of Physical Science and Technology, Southwest University, Chongqing 400715, China}

\author{Zu-Xing Yang}
\email{zuxing.yang@riken.jp}
\affiliation{RIKEN Nishina Center, Wako, Saitama 351-0198, Japan}

\author{Peng Yin}
\email{yinpeng@impcas.ac.cn}
\affiliation{Department of Physics and Astronomy, Iowa State University, Ames, IA 50011, USA}
\affiliation{Institute of Modern Physics, Chinese Academy of Sciences, Lanzhou 730000, China}

\author{Peng-Hui Chen}
\affiliation{College of Physics Science and Technology, Yangzhou University, Yangzhou, Jiangsu 225002, China}

\author{Jian-Min Dong}
\affiliation{Institute of Modern Physics, Chinese Academy of Sciences, Lanzhou 730000, China}

\author{Zhi-Pan Li}
\affiliation{School of Physical Science and Technology, Southwest University, Chongqing 400715, China}

\author{Haozhao Liang}
\affiliation{Department of Physics, Graduate School of Science, The University of Tokyo, Tokyo 113-0033, Japan}
\affiliation{RIKEN iTHEMS, Wako 351-0198, Japan}

\begin{abstract}
We adapt the local density approximation to add the high-momentum tails (HMTs) to finite nuclei's Slater-determinant momentum distributions.
The HMTs are extracted by the extended Brueckner-Hartree-Fock (EBHF) method or by the lowest order cluster approximation.
With a correction factor being added to EBHF, it is sufficiently in agreement with the experimental benchmark, i.e., the high-momentum $N/Z$ ratios approximately equal to $1$, and the low-momentum $N/Z$ ratios approximately equal to $N/Z$ of the systems.
It is also found that the tensor force makes the nucleon-nucleon correlations appear more easily on the nuclear surface region and the percentage of high-momentum ($p > 300$~MeV/c) nucleons, around $17\%$--$18\%$, independent of isospin asymmetry.
\end{abstract}

\maketitle

\section{Introduction}

Recently, worldwide experiments have revealed many exciting results on the dynamical correlations of nucleons \cite{Rohe2004Phys.Rev.Lett.93.182501, Onderwater1998Phys.Rev.Lett.81.22132216, Starink2000Phys.Lett.B474.3340,Piasetzky2006Phys.Rev.Lett.97.162504, Subedi2008Science320.14761478, Hen2014Science346.614617,Shneor2007Phys.Rev.Lett.99.072501,Duer2018Nature560.617621}.
It was surprising to observe that the proton-neutron ($pn$) short-range correlations (SRCs) in nuclei are much stronger than the proton-proton ($pp$) and neutron-neutron ($nn$) correlations by a factor of about $20$, for the internal momenta of $250$--$600$~MeV/c, where the tensor forces dominate the nucleon-nucleon interactions \cite{Piasetzky2006Phys.Rev.Lett.97.162504, Subedi2008Science320.14761478, Hen2014Science346.614617}.
The two-nucleon knockout experiment demonstrated that these $pn$-dominated correlated pairs are formed with large relative momenta and small center-of-mass momenta \cite{Shneor2007Phys.Rev.Lett.99.072501}.
Recently, the $(e,e'p)$ and $(e,e'n)$ quasi-elastic knockout event-sampling experiments have further displayed that some nucleons in nuclei form close-proximity neutron-proton pairs with high nucleon momentum at different isospin-asymmetries \cite{Duer2018Nature560.617621}.
The SRC quenches the neutron superfluidity and neutrino emissivity of neutron stars, and hence visibly affects the neutron star cooling \cite{Dong2013Phys.Rev.C87.062801,Dong2016Astrophys.J.817.6}.
Moreover, the quasi-free $\alpha$ cluster–knockout reactions showed a direct experimental evidence for forming $\alpha$ clusters at the surface of neutron-rich Sn isotopes \cite{Tanaka2021Science371.260264}.

For these new experimental discoveries, theorists are trying to provide a self-consistent and reliable explanation.
Various theoretical methods have been employed to calculate the nucleon-nucleon correlations in nuclear matter, such as the correlated basis functions \cite{Fantoni1984Nucl.Phys.A427.473492, Benhar1989Nucl.Phys.A505.267299, Benhar1990Phys.Rev.C41.R24R27}, the quantum Monte Carlo method \cite{Gezerlis2010Phys.Rev.C81.025803}, the self-consistent Green's function (SCGF) \cite{Dewulf2002Phys.Rev.C65.054316, Dewulf2003Phys.Rev.Lett.90.152501, Frick2005Phys.Rev.C71.014313, Rios2009Phys.Rev.C79.025802,Rios2009Phys.Rev.C79.064308}, the in-medium T-matrix method \cite{Bozek1999Phys.Rev.C59.26192626,Bozek2002Phys.Rev.C65.054306,Soma2008Phys.Rev.C78.054003}, and the Brueckner-Hartree-Fock (BHF) method \cite{Sartor1980Phys.Rev.C21.15461567, Yin2017Chin.Phys.C41.114102,Yin2013Phys.Rev.C87.014314,Grange1987Nucl.Phys.A473.365393, Jaminon1990Phys.Rev.C41.697705, Baldo1991Nucl.Phys.A530.135148, Mahaux1993Nucl.Phys.A553.515518, Hassaneen2004Phys.Rev.C70.054308}.
In particular, in Ref.~\cite{Yang2019Phys.Rev.C100.054325} the isospin- and density-dependent momentum distribution calculated by extended Brueckner-Hartree-Fock (EBHF) has been parameterized.
For finite nuclei, the local density approximation (LDA) based on the results of the lowest order cluster (LOC) approximation \cite{Stringari1990Nucl.Phys.A516.3340, Flynn1984Nucl.Phys.A427.253277, Fantoni1984Nucl.Phys.A427.473492, Gaidarov2009Phys.Rev.C80.054305} and the light-front dynamics method \cite{Gaidarov2009Phys.Rev.C80.054305} have been utilized to describe the momentum distributions with initial success. 
However, these methods cannot adequately explain the existence of the high-momentum $pn$-dominated close-proximity correlated pairs \cite{Duer2018Nature560.617621}.
Therefore, a phenomenological (i.e., experiment-based) $pn$-dominance model \cite{Hen2014Science346.614617, Sargsian2014Phys.Rev.C89.034305}, which uses a mean-field momentum distribution at low momentum ($k < k_f$) and a scaled deuteron-like high-momentum tail, has been developed.

In this work, we employ the LDA method to include the high momentum tails (HMTs) in finite nuclei as a significant correction to the Slater determinant momentum distributions.
This paper is organized as follows.
The theoretical approaches, including the EBHF theory and the LOC approximation, are briefly reviewed in Sec.~\ref{Sect:II}.
In Sec.~\ref{Sect:III}, the momentum properties with the two methods are compared, and a modification to the EBHF is proposed as a new scheme. 
In Sec.~\ref{Sect:IV}, we employ the modified model to study the SRC effects on selected nuclei and compare the results with the available experimental data. 
Finally, a summary is given in Sec.~\ref{Sect:V}.

\section{Theoretical Framework}\label{Sect:II}

In nuclear matter, dynamical correlations modify the occupation probability of nucleon from that in the Fermi gas model.
At zero temperature, this process can be characterized as \cite{Stringari1990Nucl.Phys.A516.3340}
\begin{equation}
n^\tau_{\rm NM}(k;\rho,\delta) = \Theta(k^\tau_f - k) + \delta n^\tau_{\rm NM}(k;\rho,\delta).
\label{eq:nnmk}
\end{equation}
Here, the nuclear matter is characterized by its total density $\rho = \rho_n+\rho_p$ and isospin asymmetry $\delta = (\rho_n-\rho_p)/(\rho_n+\rho_p)$.
The corresponding Fermi momenta read $k^\tau_f = (3 \pi^2 \rho_\tau)^{1/3}$ for neutrons and protons ($\tau = n, p$), respectively.
The $n^\tau_{\rm NM}(k)$, which is dimensionless and satisfies $0 \le n^\tau_{\rm NM}(k) \le 1$, is the correlated momentum distribution in nuclear matter, $\Theta(k^\tau_f - k)$ denotes the occupation probability of the independent-particle model. And $\delta n^\tau_{\rm NM}(k)$ is the correction caused by the dynamical correlations.
By definition, $\int \delta n^\tau_{\rm NM}(k)\, d^3k = 0$ to ensure the conservation of particle numbers.

For a finite nucleus, the momentum distribution can also be written as a sum of the single-particle contribution and the correlation effect, i.e.,
\begin{equation}
  n^\tau_{A}(k) = n^\tau_{\rm SD}(k) + \delta n^\tau_{A}(k),
\label{eq:nak}
\end{equation}
where
\begin{equation}
  n^\tau_{\rm SD}(k) = \sum_{\alpha\in \tau} v^2_\alpha \phi^*_\alpha(k) \phi_\alpha(k)
\end{equation}
is the Slater-determinant momentum distribution generated by the single-particle wave functions $\phi_\alpha(k)$ written in the momentum-space representation, with the corresponding occupation probabilities $v^2_\alpha$.
The subscript $A$ labels the nuclide and $\alpha$ labels the single-particle quantum numbers.
The $\delta n^\tau_{A}(k)$ corresponds to the dynamical correlation, satisfying $\int \delta n^\tau_{A}(k)\, d^3 k = 0$ due to the particle number conservation. 
Note that, in the whole paper, the momentum distributions in finite nuclei are normalized to the particle numbers, i.e.,
\begin{equation}
  \int n^\tau_{A}(k)\, d^3k = \mathcal{N}^\tau,
\end{equation}
with $\mathcal{N}^\tau$ the number of protons ($Z$) or neutrons ($N$).
The unit of $n^\tau_{A}(k)$ is fm$^3$.

In this paper, $n_{\rm SD}(k)$ is calculated by the self-consistent Skyrme Hartree-Fock (SHF) model with Bardeen-Cooper-Schrieffer (BCS) pairing, by using the SkM* interaction \cite{Bartel1982Nucl.Phys.A386.79100} and adopting the spherical symmetry.

Based on LDA, one can obtain the $\delta n_{A}^\tau(k)$ from the superposition of $\delta n_\text{NM}^\tau(k)$ at different densities,
\begin{equation}
\delta n_{A}^\tau(k) =\frac{\int \lambda^\tau(r) \, \delta n_{\rm NM}^\tau(k; \rho(r), \delta(r)) \rho_\tau(r)\, d^3 r}{\mathcal{N}^\tau},
\label{eq:dnak}
\end{equation}
where $\rho_\tau(r)$ is the Slater-determinant density distribution of proton or neutron, $\rho(r)$ and $\delta(r)$ are the corresponding local total density and isospin asymmetry, respectively.
The normalization factor $\lambda^\tau(r) = \mathcal{N}^\tau/[4\pi^3 \rho_\tau(r)]$ here takes care of the differences in the units and normalization conditions between $n^\tau_{\rm NM}(k)$ and $n^\tau_A(k)$.
Combining Eqs.~(\ref{eq:nnmk}), (\ref{eq:nak}), and (\ref{eq:dnak}), one can obtain the correlated momentum distributions of a finite nucleus.
To this end, the EBHF method or the LOC approximation is adopted to determine the only unknown quantity $\delta n^\tau_{\rm NM}(k)$.

\subsection{Extended Brueckner-Hartree-Fock Method}

The Brueckner-Hartree-Fock method is one of the widely used \textit{ab initio} approaches for inverstigating the properties of nuclear matter.
In Refs.~\cite{Zuo2002Eur.Phys.J.A14.469475, Grange1989Phys.Rev.C40.10401060}, the BHF model with the realistic Argonne V18 \cite{Wiringa1995Phys.Rev.C51.3851} two-body interaction was extended to include the microscopic three-body force, and it is called the EBHF method.
For the details of the EBHF method, one can refer to Refs.~\cite{Zuo1999Phys.Rev.C60.024605, Grange1989Phys.Rev.C40.10401060, Zuo2002Eur.Phys.J.A14.469475, Zuo2002Nucl.Phys.A706.418430}.

In this scheme, the realistic nuclear force is converted into the effective interaction G-matrix of the Bethe-Brueckner-Goldstone theory by a self-consistent solution of the Bethe-Goldstone equation.
This G-matrix, which includes all ladder diagrams of nucleon-nucleon interactions and  embodies the tensor correlations and SRCs, can be used to compute the mass operator $M(k,\omega)$ \cite{Yang2019Phys.Rev.C100.054325}.
The mass operator $M(k,\omega)$ allows us to write down the Green's function in the energy-momentum representation,
\begin{equation}
\mathcal{G}(k, \omega)=\frac{1}{\omega-\frac{k^{2}}{2 m}-M(k, \omega)} .
\end{equation}
Futhermore, the spectral function $S(k, \omega)$, which describes the probability density of removing a particle with momentum $k$ from a target nuclear system and leaving a final system with excitation energy $\omega$, is thus given by
\begin{equation}
S(k, \omega)=\frac{i}{2 \pi}\left[\mathcal{G}(k, \omega)-\mathcal{G}(k, \omega)^{*}\right],
\end{equation}
with the sum rule $\int_{-\infty}^{\infty} S(k, \omega)\, d \omega = 1$.
Finally, one can obtain the momentum distributions using the spectral function by
\begin{equation}
n_{\rm NM}(k)=\int_{-\infty}^{\varepsilon_{f}} S(k, \omega)\, d \omega,
\end{equation}
where the Fermi energy $\varepsilon_{f}$ satisfies  the on-shell condition $\varepsilon_{f}=k_{f}^{2} / 2 m+\text{Re} ~M\left(k, \varepsilon_{f}\right)$.

\begin{widetext}
As a result, the momentum distributions in nuclear matter can be parameterized as \cite{Yang2019Phys.Rev.C100.054325}
\begin{equation}
n^\tau_{\rm NM}(k;\rho,\delta)=\left\{\begin{array}{lll}
\displaystyle\frac{1-\chi^\tau}{0.9546}\left[1.0033-0.0288 \frac{k}{k_{f}^{\tau}}-0.0905(\frac{k}{k_{f}^{\tau}})^{7}\right], & & 
\text {for } k \leq k_{f}^{\tau}, \\
\\
\displaystyle\frac{\chi^\tau}{2.9537}\left[3.548\, e^{-1.799 {k}/{k_{f}^\tau}}+52.2\, e^{-4.2766({k}/{k_{f}^{\tau}})^{2}}\right], & &
\text {for } k > k_{f}^{\tau},
\end{array}\right.
\end{equation}
where
\begin{equation}
\chi^\tau(\rho, \delta) = 0.1669\left[1+\lambda(0.1407 \frac{\rho}{\rho_{0}}-0.7296) \delta\right]
\left[1+2.448\, e^{-4.1854{\rho}/{\rho_{0}}} +0.1382(\frac{\rho}{\rho_{0}})^{1.5}\right],
\end{equation}
with $\rho_0 = 0.17$~fm$^{-3}$ being saturation density and $\lambda = 1$ and $-1$ corresponding to neutron and proton, respectively.
The physical meaning of $\chi^\tau$ is the percentage of nucleons leaping above the Fermi sea due to the correlations, i.e., $\chi^\tau=\frac{1}{\rho_\tau\pi^{2}} \int_{k_{f}^\tau}^{\infty} n^\tau_{\rm NM}(k) k^{2}\, d k$.
See Ref.~\cite{Yang2019Phys.Rev.C100.054325} for details.
\end{widetext}

\subsection{Lowest Order Cluster Approximation}

In contrast, the phenomenological LOC approximation developed in Ref.~\cite{Flynn1984Nucl.Phys.A427.253277} has also been used to evaluate the correlated term.
Eqs.~(\ref{eq:nak}) and (\ref{eq:dnak}) can be reduced to
\begin{equation}
n_{A}^\tau(k) = n_{\rm SD}^\tau(k) +\frac{1}{4\pi^3}\int \delta n_{A}^\tau(k;\rho(r), \delta(r))\, d^3{r}.
\label{eq:nakt}
\end{equation}
Choosing a correlation function $f(r) = 1-e^{-\beta^2 r^2}$, the $\delta n_{A}^\tau(k;\rho(r), \delta(r))$ can be given by the LOC approximation,
\begin{equation}
\begin{aligned}
\delta n_{A}^\tau(k;\rho(r), \delta(r)) =&~\left[Y(k, 8)-k_{\rm dir}\right] \Theta\left(k_f^\tau(r)-k\right) \\
&~ +8\left\{k_{\rm dir} Y(k, 2)-[Y(k, 4)]^{2}\right\},
\end{aligned}
\end{equation} 
where
\begin{equation}
\begin{aligned}
c_{\mu}^{-1} Y(k, \mu) =&~ \frac{e^{-\tilde{k}_{+}^{2}}-e^{-\tilde{k}_{-}^{2}}}{2 \tilde{k}}+\int_{0}^{\tilde{k}_{+}} e^{-y^{2}}\, dy \\
&~ +\operatorname{sgn}(\tilde{k}_{-}) \int_{0}^{\left|\tilde{k}_{-}\right|} e^{-y^{2}}\, dy,
\end{aligned}
\end{equation}
with $c_{\mu}=\frac{1}{8 \sqrt{\pi}}\left(\frac{\mu}{2}\right)^{3 / 2}$, $\tilde{k}=\frac{k}{\beta \sqrt{\mu}}$, $\tilde{k}_{\pm}=\frac{k_{f} \pm k}{\beta \sqrt{\mu}}$, and $\operatorname{sgn}(x)=\frac{x}{|x|}$. 
The quantity $k_{\rm dir}$ is the direct part of the Jastrow wound parameter, written as
\begin{equation}
\begin{aligned}
k_{\rm dir} &=\frac{2{k_{f}^{\tau}}^{3}}{3 \pi^{2}} \int[f(r)-1]^{2}\, d^3{r} \\
&=\frac{1}{3 \sqrt{2 \pi}}\left(\frac{k_{f}^{\tau}}{\beta}\right)^{3}.
\end{aligned}
\end{equation}
A reasonable range of values of $k_{\rm dir}$ for nuclear matter is 0.1-0.3 \cite{Stringari1990Nucl.Phys.A516.3340, Flynn1984Nucl.Phys.A427.253277}, where a larger value corresponds to a stronger correlation effect. 
For more microscopic calculations and evaluations, please refer to Refs.~\cite{Flynn1984Nucl.Phys.A427.253277, Pandharipande1979Rev.Mod.Phys.51.821861}. 

The tensor correlations and SRCs can be considered within the above two frameworks.
In Ref.~\cite{Gaidarov2009Phys.Rev.C80.054305},
Gaidarov \textit{et al.} added the BCS correlations to the mean-field term $n_{\rm SD}(k)$ and discussed the effect of BCS correlations.
In this study, we also take into account the BCS effect.

\section{Comparison and phenomenological correction}\label{Sect:III}

\begin{figure}
\includegraphics[width=9 cm]{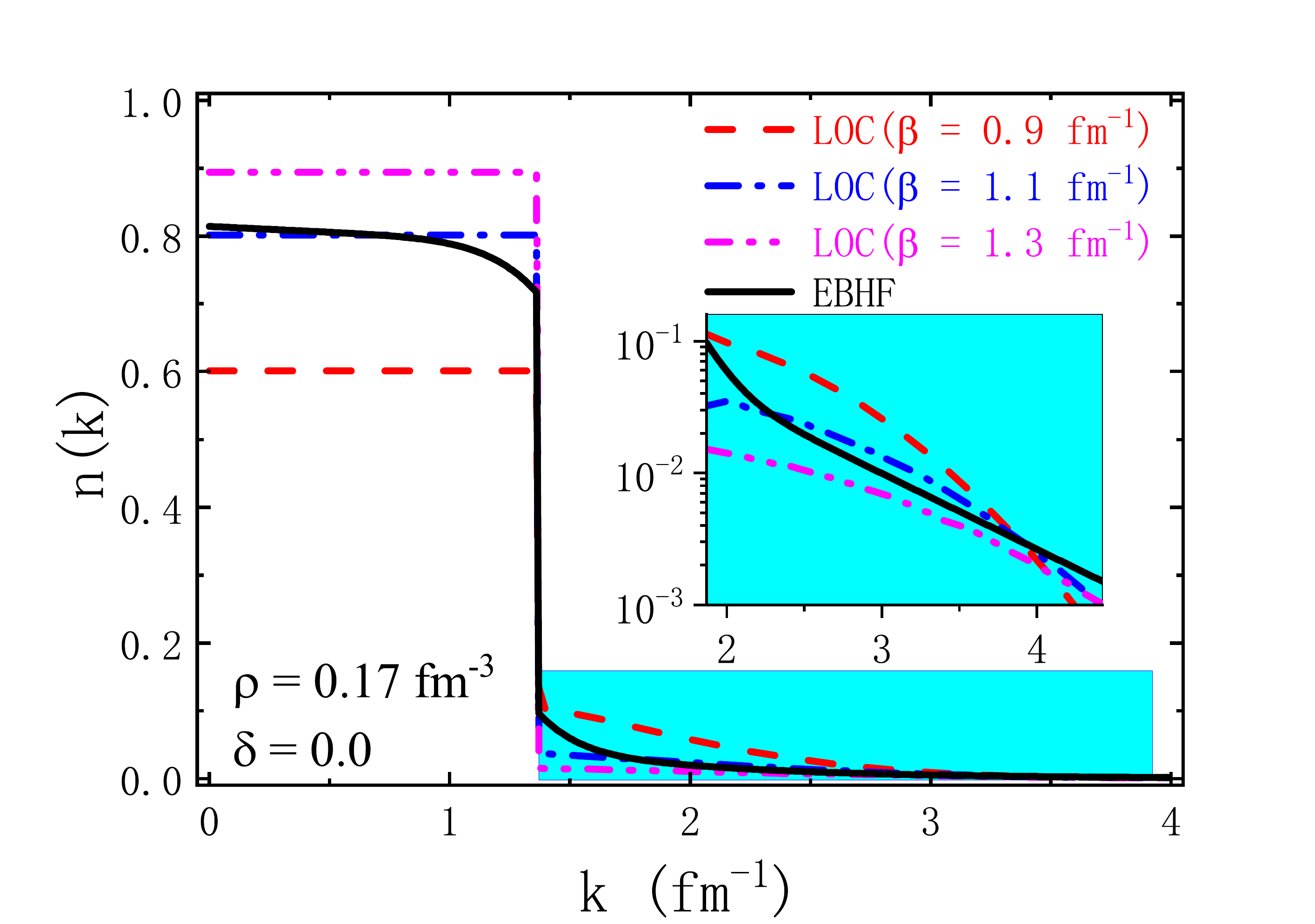}
\caption{\label{fig1} (Color online) Momentum distributions of symmetric nuclear matter at saturation density calculated by the EBHF method and the LOC approximation with the correlation factor $\beta = 0.9, 1.1, 1.3$~fm$^{-1}$. Inset: HMTs zoomed with a logarithmic scale.}
\end{figure}

We start with the comparison between the two methods mentioned in the previous section.
The early applications of the LOC approximation focused on the isospin-independent case \cite{Stringari1990Nucl.Phys.A516.3340, Flynn1984Nucl.Phys.A427.253277}. 
Thus, we first show the momentum distributions of the symmetric nuclear matter at the saturation density calculated by the EBHF method and the LOC approximation in Fig.~\ref{fig1}.

On the one hand, the results of LOC are obtained by taking the correlation factor $\beta = 0.9$, $1.1$, and $1.3$~fm$^{-1}$ as examples.
It is evident that the value of $\beta$ does not significantly affect the shape of momentum distribution below and above the Fermi surface, but it affects the proportion of HMTs in the system.
Specifically, the proportion of HMTs reaches $40\%$, $20\%$, and $10\%$ with $\beta = 0.9$, $1.1$, and $1.3$~fm$^{-1}$, respectively.
Since the proportion of the high-momentum nucleons is experimentally considered to be about $20\%$ \cite{Hen2014Science346.614617,Subedi2008Science320.14761478}, hereafter we take $\beta = 1.1$~fm$^{-1} ~(k_{dir} = 0.23)$ for further discussions.

On the other hand, it is remarkable that the result obtained by the EBHF method shows essentially different features, compared to the LOC results.
First, there is a significant depletion in the momentum distribution by EBHF just below the Fermi surface, while such a feature is missing by LOC.
Moreover, as emphasized in the inset of Fig.~\ref{fig1}, the LOC results exhibit a nearly linear relation between $n(k)$ and $k$ in the high-momentum tail, which is, however, consistent with neither Tan's relation \cite{Hen2015Phys.Rev.C92.045205, Tan2008Ann.Phys.323.29522970, Tan2008Ann.Phys.323.29712986, Tan2008Ann.Phys.323.29872990, Stewart2010Phys.Rev.Lett.104.235301} nor other theoretical calculations.
In contrast, the EBHF result exhibits a nearly exponential relation $n(k)\propto \exp(-k)$ in the tail, which is consistent with the calculation by the SCGF method \cite{Rios2009Phys.Rev.C79.064308}.
Moreover, for different interactions, the depletions of the Fermi sea would differ from each other \cite{Rios2014Phys.Rev.C89.044303, Rios2009Phys.Rev.C79.064308}.
Nevertheless, below the saturation density, such depletions show the same trend, which is always decreasing with increasing density \cite{Li2016Phys.Rev.C94.024322}.
At low densities, the momentum distribution exhibits a weak sensitivity to nuclear forces.
Taking the meson-exchange CD Bonn potential as an example, it preforms similarly to Argonne V18 below the normal nuclear density in the SCGF method \cite{Rios2009Phys.Rev.C79.064308, Rios2009Phys.Rev.C79.025802}.

\begin{figure}
\includegraphics[width=9 cm]{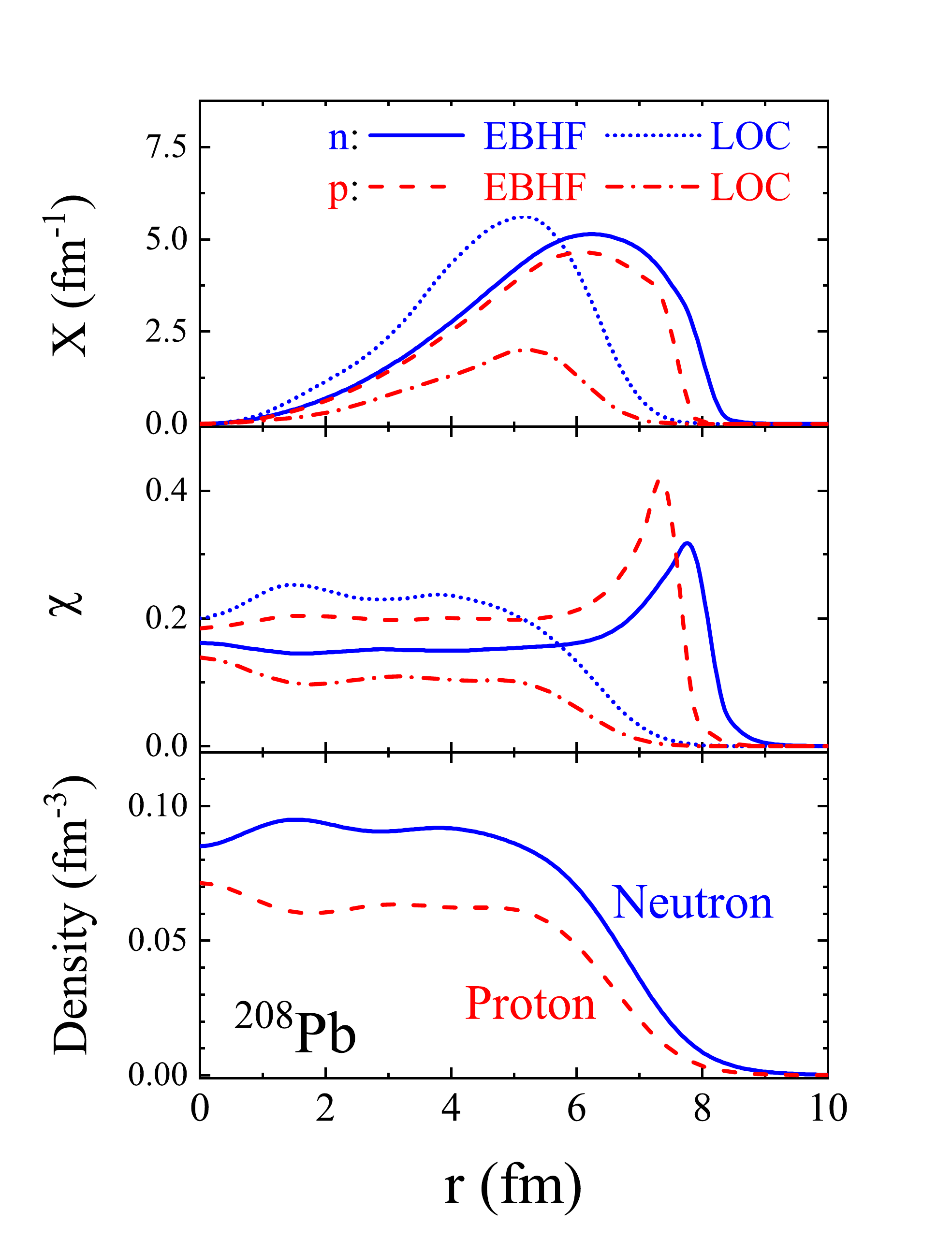}
\caption{\label{fig2} (Color online) Lower panel: Density distributions of neutrons and protons of $^{208}$Pb calculated by SHF with SkM*. Middle panel: Correlation strengths $\chi$ for neutrons and protons as a function of radius calculated by the EBHF method and the LOC approximation. Upper panel: Number densities of correlated neutrons and protons $X$ as a function of radius.}
\end{figure}

With the help of LDA shown in Eqs.~(\ref{eq:nak}) and (\ref{eq:dnak}), the two mentioned methods can be applied to finite nuclei.
We take $\rm ^{208}Pb$ as an example and plot the corresponding results in Fig.~\ref{fig2}.
The lower panel of Fig.~\ref{fig2} is the density distributions of neutrons and protons of $\rm ^{208}Pb$ calculated by the SHF method with the effective interaction SkM*; the middle panel shows the correlation strengths $\chi$ as a function of radius $r$; and the upper panel manifests the number densities of correlated protons and neutrons $X$ as a function of radius $r$, i.e.,
\begin{equation}
  X^\tau(r) = 4 \pi r ^2 \rho^\tau(r) \chi^\tau(r).
\label{eq:Xtau}
\end{equation}

One can see from the middle panel of Fig.~\ref{fig2} that the EBHF result exhibits stronger correlations on the nuclear surface than in the interior region, while the LOC result exhibits the correlation strengths with similar shapes of the corresponding density distributions.
This is due to the fact that EBHF takes into account the tensor correlation, compared to LOC.
Below the saturation density, the depletion of the Fermi sea becomes stronger with decreasing density, which mainly results from the increasing effect of the tensor correlation \cite{Yang2019Phys.Rev.C100.054325}.
This illustrates that the tensor force dominates the nucleon-nucleon correlation on the nuclear surface \cite{Rios2014Phys.Rev.C89.044303},  which may also responsible for the formation of $\alpha$-cluster on the surface of the Sn isotopes mentioned in a recent quasi-free $\alpha$ cluster-knockout experiment \cite{Tanaka2021Science371.260264}.
Additionally, inclusion of the three-body force leads to an overall enhancement of the depletion of the neutron and proton Fermi seas within EBHF.
However, such an effect on the neutron and proton momentum distributions turns out to be negligibly weak around and below the normal nuclear density \cite{Yin2013Phys.Rev.C87.014314, Zuo2002Nucl.Phys.A706.418430}. 

In the upper panel of Fig.~\ref{fig2}, the areas enclosed by the curves and the horizontal axis qualitatively reflect the numbers of correlated nucleons.
One problem with the isospin dependence of the LOC approximation is the apparently different numbers of correlated protons and neutrons, which is essentially inconsistent with the experimental discoveries.
In contrast, the EBHF result shows similar numbers of correlated protons and neutrons, i.e.,
\begin{equation}
\int_{k_f^n}^{\infty} n^n(k) k^2\, dk \approx \int_{k_f^p}^{\infty} n^p(k) k^2\, dk.
\label{eq:knapkp}
\end{equation}
However, in the $pn$-dominance picture \cite{Hen2014Science346.614617, Yong2018Phys.Lett.B776.447450, Sargsian2014Phys.Rev.C89.034305}, the numbers of neutrons and protons in the HMTs should approximately be the same, i.e.,
\begin{equation}
\int_{k_{\rm high}}^{\infty} n^n(k) k^2\, dk \approx \int_{k_{\rm high}}^{\infty} n^p(k) k^2\, dk.
\label{eq:khigh}
\end{equation}
with the $k_{\rm high} = 300~\mathrm{MeV/c}$ based on the recent experimental observations.
The discrepancy that exists between Eqs.~(\ref{eq:knapkp}) and (\ref{eq:khigh}) is caused by the Fermi momentum gap between protons and neutrons due to the difference in their densities.

A proton momentum gap was proposed to eliminate the above contradiction in the studies of heavy-ion collisions \cite{Yong2018Phys.Lett.B776.447450, Yong2022Phys.Rev.C105.l011601}.
Such a proposal partially succeeded at the cost of the continuity of the momentum distribution.
In the present work, we introduce an isospin-dependent correction factor (CF) $\xi$ for the correlation term $\delta n_{\rm NM}(k)$ given by EBHF as a softer ansatz:
\begin{equation}
\begin{aligned}
\delta n_{\rm NM}^{\rm EBHF}(k) &\rightarrow \xi\, \delta n_{\rm NM}^{\rm EBHF}(k), \\
\xi &= {(\frac{N}{Z})}^{-\lambda/2},
\end{aligned}
\end{equation}
where $\lambda = 1$ and $-1$ corresponds to neutron and proton, respectively.
By definition, for symmetric nuclei, $\xi$ shows basically no effects.
For asymmetric nuclei, $\xi$ makes Eq.~(\ref{eq:khigh}) theoretically valid.
A natural speculation is that the existence of $\xi$ might be related to the kinetic-energy part of the symmetry energy.
In short, based on LDA, we adopt three schemes to calculate the correlation terms of the momentum distributions in finite nuclei---LOC, EBHF, and EBHF with CF---which will be further compared in the next section.

\section{Results and Discussion}\label{Sect:IV}

\begin{figure}
\includegraphics[width=9.5 cm]{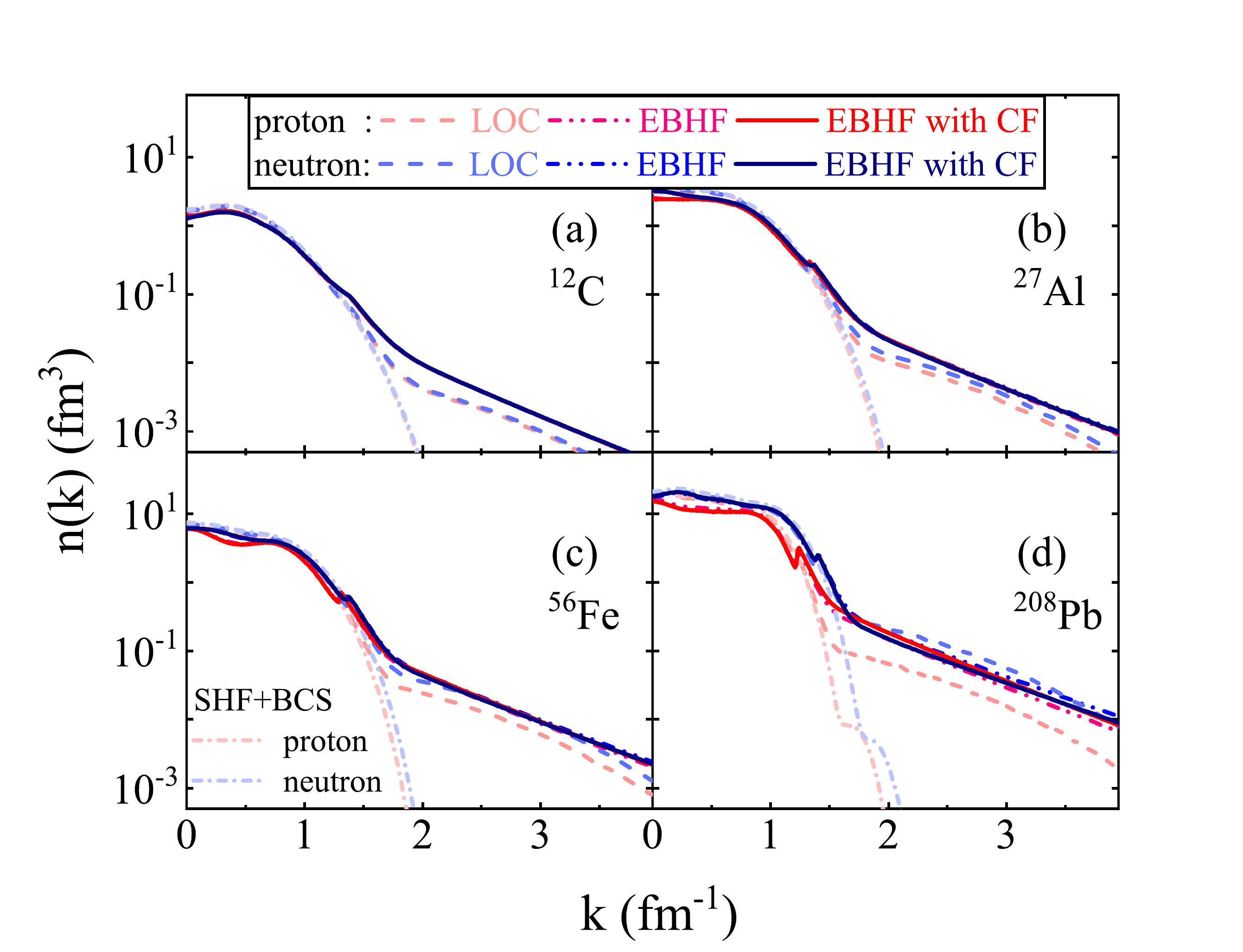}
\caption{\label{fig4} (Color online) Dynamical correlation effects of momentum distributions in (a) $\rm ^{12}C$, (b) $\rm ^{27}Al$, (c) $\rm ^{56}Fe$, and (d) $\rm ^{208}Pb$ calculated by LDA.
The correlation terms are obtained by the EBHF method with and without CF and by the LOC approximation.}
\end{figure}

In the applications for finite nuclei, we first take $^{12}$C, $^{27}$Al, $^{56}$Fe, and $^{208}$Pb as examples. 
The dynamical correlation effects of momentum distributions are shown in Fig.~\ref{fig4}, where the correlation terms are obtained by the EBHF method with and without CF and by the LOC approximation.
In addition, the momentum distributions calculated with SHF+BCS, without the tensor correlation and SRCs effects, are also plotted with the light red curves for proton and light blue curves for neutron for comparison.
Calculations with SHF+BCS indicate that there are still about $10\%$ nucleons above the Fermi surface, where $2\%$ stems from the BCS correlation as suggested in Ref.~\cite{Gaidarov2009Phys.Rev.C80.054305} taking $^{84}$Kr as an example.
It is seen that the high-momentum nucleons obtained by the LOC approximation are less than those obtained by the EBHF method.
According to Fig.~\ref{fig2}, the EBHF method generates more high-momentum nucleons on the surface of a nucleus.

Comparing the four panels of Fig.~\ref{fig4}, the difference between the numbers of high-momentum protons and neutrons obtained by LOC increases gradually with increasing isospin asymmetry.
Such a difference is alleviated in the EBHF framework and eliminated by that with CF.
It is clear that the HMTs of protons and neutrons calculated by EBHF with CF overlap precisely in the domain ($k>1.52$~fm$^{-1}$, i.e., $p>300$~MeV/c) as expected, and no longer depend on the isospin asymmetry.

\begin{figure}
\includegraphics[width=9 cm]{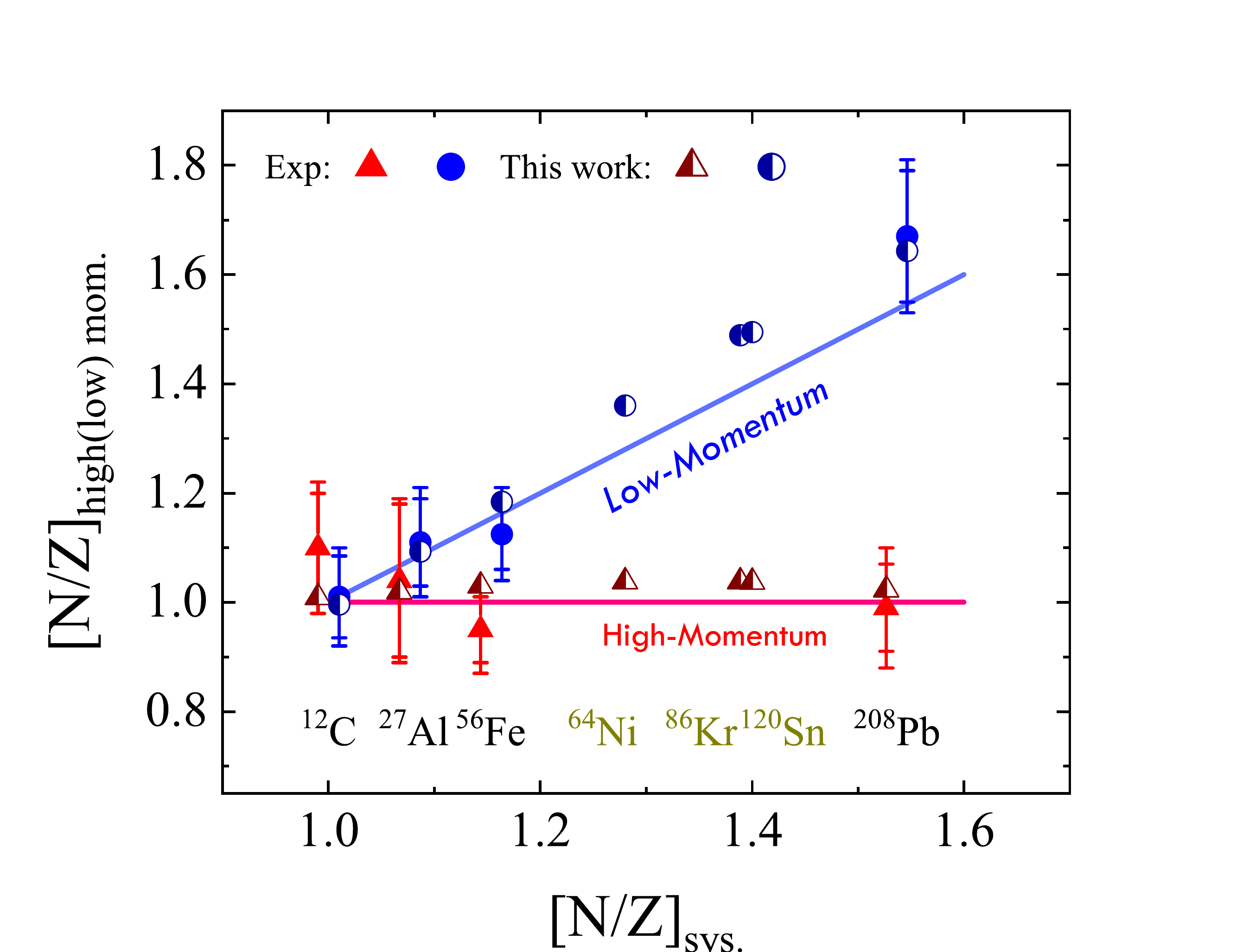}
\caption{\label{fig5} (Color online) 
The $N/Z$ ratio for the low-momentum ($p < 250$~MeV/c) nucleons and the high-momentum ($p > 300$~MeV/c) nucleons, shown by circles and triangles, respectively.
The calculated results by EBHF with CF are shown with the half-filled symbols.
The experimental data \cite{Duer2018Nature560.617621} are shown with the filled symbols, where the inner (outer) error bars correspond to the statistical (statistical and systematic) uncertainties.
The lines with $N/Z$ ratio equals to $1$ and $N/Z$ are also drawn for guiding eyes.
}
\end{figure}

\begin{figure}
\includegraphics[width=9 cm]{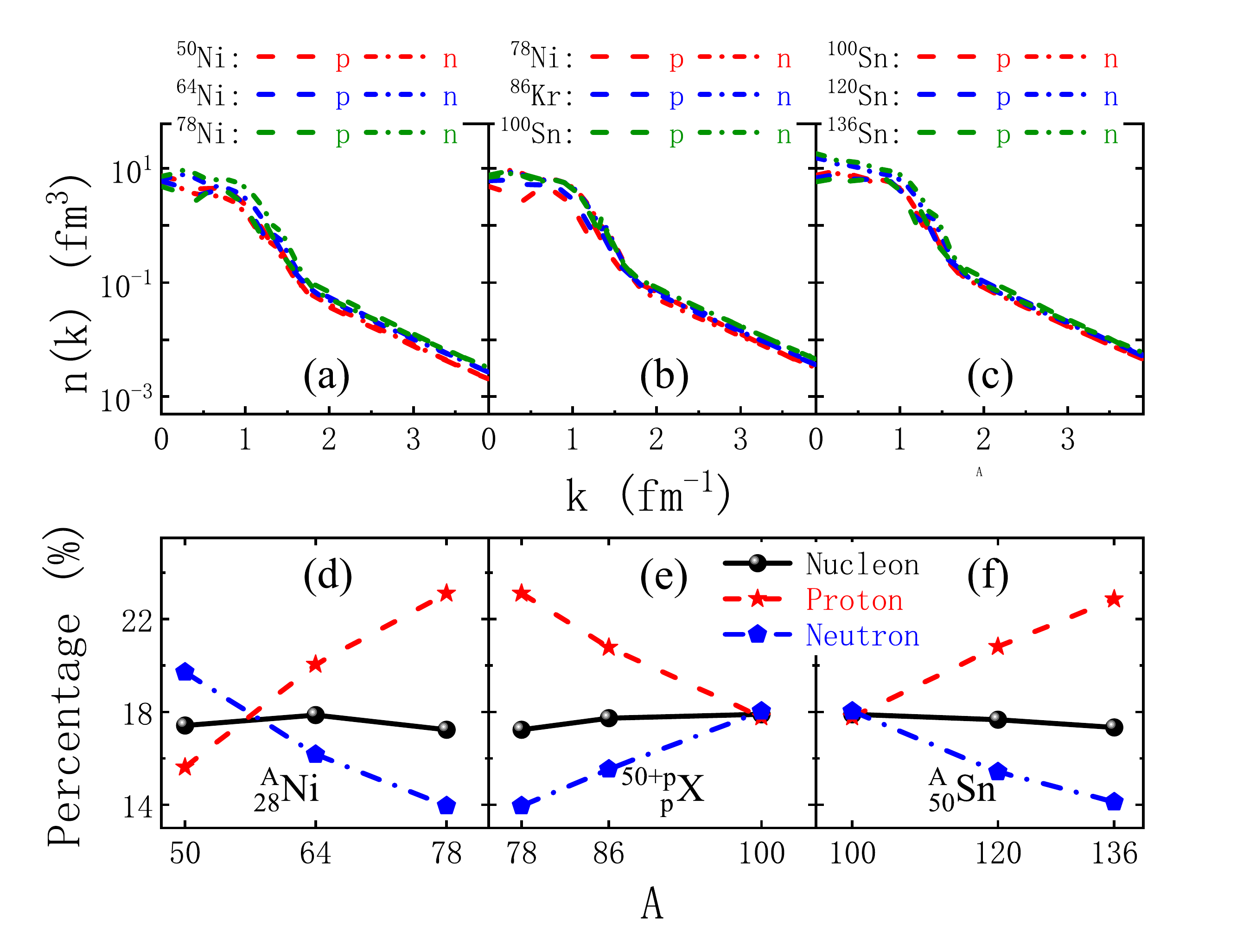}
\caption{\label{fig6} (Color online) Upper panels: Momentum distributions for (a) $^{A}_{28}$Ni ($A = 50, 64, 78$), (b) $^{50+p}_{~~~~p}$X ($p = 28, 36, 50$), and (c) $^{A}_{50}$Sn ($A = 100, 120, 136$) calculated by EBHF with CF.
Lower panels: The corresponding percentages of the high-momentum nucleons as a function of nucleon number for (d) $^{A}_{28}$Ni, (e) $^{50+p}_{~~~~p}$X, and (f) $^{A}_{50}$Sn.}
\end{figure}

As a step further, the $N/Z$ ratios for the low-momentum nucleons and the high-momentum nucleons calculated by EBHF with CF are shown in Fig.~\ref{fig5}, together with the experimental data.
We adopt the experimental definition of low momentum as $p < 250$~MeV/c and high momentum as $p >300$~MeV/c, by which one can integrate the corresponding interval of the distributions in Fig.~\ref{fig4} and gain the $N/Z$ ratios in Fig.~\ref{fig5}.
The fully-filled red triangles and blue circles represent the experimental data \cite{Duer2018Nature560.617621}, and the half-filled symbols are the predictions by EBHF with CF.
Excellent agreements between the theoretical predictions and the experimental data are achieved, for $^{12}$C, $^{27}$Al, $^{56}$Fe, and $^{208}$Pb.
Moreover, the results satisfy the simple $N/Z$ expectations, i.e., the high-momentum $N/Z$ ratios approximately equal to $1$, while the low-momentum $N/Z$ ratios approximately equal to $N/Z$, the isospin asymmetry of the system.

In Fig.~\ref{fig5}, the corresponding predictions for the unmeasured $^{64}$Ni,  $^{86}$Kr, and $^{120}$Sn are also shown.
These results also follow the systematics of the high-momentum and low-momentum $N/Z$ ratios.

We can further investigate the nature of HMTs along the isotopic and isotonic chains.
Panels (a)--(c) of Fig.~\ref{fig6} show the momentum distributions calculated by EBHF with CF for the Ni ($Z=28$) isotopes, the $N=50$ isotones, and the Sn ($Z=50$) isotopes, respectively, while panels (d)--(f) display the corresponding percentages of the high-momentum nucleons as a function of nucleon number $A$.
It is found that the shape of HMT is always the same in panels (a)--(c).
In addition, panels (d)--(f) exhibit an important property that the proportion of high-momentum nucleons is always in the range of $17\%$--$18\%$, independent of density distribution or isospin asymmetry of a nucleus.
From (d) and (f), one can see that with the increasing number of neutrons in the system, the number of correlated protons gradually increases.
Similarly, when the number of protons in the system increases, the number of correlated neutrons also gradually increases, as shown in panel (e).
These conclusions are, in general, consistent with the experiment.
But in terms of detail, experiment also shows that the strength of HMT grows with mass number ($A$) \cite{Fomin2012Phys.Rev.Lett.108.092502}.
This suggests that further study should account for the $A$-dependence of the depletion $\chi^\tau$ or the correlation parameter $\beta$ in applying the LDA.

From the above discussions, we speculate that, for a rich-neutron nucleus, there will be more high-momentum protons and these protons will be more likely to appear on the nuclear surface, which in turn increases the probability of $\alpha$-cluster formation.
These findings may explain the quasi-free $\alpha$-cluster knockout reactions of the neutron-rich Sn isotopes from another perspective.
This effect will also inevitably cause more $\pi^+$ mesons, photons, and free protons to be generated by the peripheral heavy-ion collisions with medium- and high-energy, which can be verified by experiments.
Furthermore, since the correlated high-momentum nucleons are more likely to appear on the nuclear surface, one may conclude that the proportion of high-momentum nucleons in the deformed nuclei is higher due to a larger surface area, compared with the same nucleus with spherical shape.
This will further affect the collective flow and viscosity in relativistic heavy-ion collisions between the deformed nuclei \cite{Zhang2022Phys.Rev.Lett.128.022301}.
One would further explore and revise it by using transport models and heavy-ion collision experiments.

The present study still has some shortcomings that can be improved in the future.
The nature of correlations in terms of wave functions and nuclear forces of finite nuclei should be further considered in order to be fully self-consistent from the theoretical points of view.
In the case of nuclear matter, one should further investigate how to deduce correction factors or other relevant physical quantities from the ladder diagrams for EBHF.

\section{Summary}\label{Sect:V}

We employ the local density approximation to give a new approach to describe the momentum distributions of finite nuclei by combining the Slater-determinant momentum distributions calculated by the SHF+BCS model with the scaled high-momentum tails calculated by the EBHF model.
On this basis, we inserted a phenomenological correction factor $\xi$ in front of the correlation term $\delta n_{\rm NM}^{\rm EBHF}(k)$ in order to solve the problem that HMTs of protons and neutrons are not completely the same due to different Fermi surfaces.
We compared our results with the $(e,e'N)$ quasi-elastic knockout event experiments of $^{12}$C, $^{27}$Al, $^{56}$Fe, and $^{208}$Pb.
We evaluated the role of correction factors for the correlation terms calculated by the EBHF method and compared with those by the LOC approximation.

It can be observed that, the correction factor makes the high-momentum distributions of protons and neutrons almost identical, which is consistent with the experiments.
Contrary to the results by the LOC approximation, we find that the correlated nucleons calculated by EBHF are more likely to appear on the nuclear surface, which may be used to explain the $\alpha$ clustering effect on the surface of Sn isotopes.
In further studies, the neutron-proton ratios for high (low) momentum nucleons of $^{64}$Ni,  $^{86}$Kr, and $^{120}$Sn are predicted and are consistent with the experimental predictions, i.e., the high-momentum $N/Z$ ratio is about $1$, and the low-momentum $N/Z$ ratio approximately equals $N/Z$.
We also calculated the Ni and Sn isotopes, and $N = 50$ isotones.
It is found that the number of correlated protons (neutrons) increases when there are more neutrons (protons) in the nuclear system.
Moreover, the number of high-momentum nucleons ($p > 300$~MeV/c) is almost independent of specific nuclides and consistently account for $17\%$--$18\%$ of all nucleons.

The conclusions of the present study can be confirmed by relevant nuclear reaction experiments or astrophysical observations, which can also be utilized to constrain further the microscopic theoretical models in turn. 

\section*{Acknowledgements}

We appreciate Professor Xin-Le Shang for his valuable advice.
This work is supported by the National Natural Science Foundation of China under Grants No.~12005175,
the Fundamental Research Funds for the Central Universities under Grant No.~SWU119076,
the JSPS Grant-in-Aid for Early-Career Scientists under Grant No.~18K13549,
the JSPS Grant-in-Aid for Scientific Research (S) under Grant No.~20H05648,
the National Natural Science Foundation of China under Grants Nos. 11705240, 12105241,
and the Natural Science Foundation of Jiangsu under Grant No.~BK20210788.
This work was also partially supported by the CUSTIPEN (China-U.S. Theory Institute for Physics with Exotic Nuclei) funded by the U.S. Department of Energy, Office of Science under Grant No.~DE-SC0009971, the RIKEN Pioneering Project: Evolution of Matter in the Universe.

\bibliographystyle{apsrev4-1}
\bibliography{ref}

\end{document}